\PassOptionsToPackage{table}{xcolor}

\documentclass[10pt,conference]{IEEEtran}

\usepackage{microtype}
\usepackage{xcolor}
\usepackage{algorithm}
\usepackage{algpseudocode}
\usepackage{xspace}
\usepackage{listings}
\usepackage{color}
\usepackage{multirow}
\usepackage{dblfloatfix}
\usepackage{lscape} 
\usepackage{soul}
\usepackage{xparse}
\usepackage{placeins}
\usepackage{caption} 
\usepackage{subcaption}
\usepackage{enumitem}
\usepackage{array, makecell}
\usepackage{wrapfig}
\usepackage{colortbl}
\usepackage{tabularx} 
\usepackage{booktabs}
\usepackage{courier}
\usepackage{ragged2e}
\usepackage{amsthm}
\usepackage{amssymb}
\usepackage{hyperref}
\usepackage{comment}
\usepackage{pdflscape}
\usepackage{afterpage}
\usepackage{fontawesome}
\usepackage{pifont} 
\usepackage{lscape}
\usepackage{svg}
\usepackage{pgfplots}
\usepackage[all]{nowidow}

\definecolor{successcolor}{RGB}{85, 157, 12}
\definecolor{failurecolor}{RGB}{209, 24, 25}

\setlist[itemize]{noitemsep, topsep=4pt}
\definecolor{lightgray}{HTML}{f6f6f6}
\definecolor{darkgray}{rgb}{.4,.4,.4}
\definecolor{darkblue}{HTML}{1b4db3}
\definecolor{brickred}{HTML}{b04f4f}
\definecolor{purple}{rgb}{0.65, 0.12, 0.82}
\definecolor{diffadd}{HTML}{288f26}
\definecolor{diffrmbg}{HTML}{ffebe9}
\definecolor{diffaddbg}{HTML}{e6ffeb}
\definecolor{diffremove}{HTML}{de4f54}
\definecolor{carrotorange}{rgb}{0.8, 0.33, 0.0}
\definecolor{highlight}{HTML}{fefbc2}
\lstdefinelanguage{JavaScript}{
  keywords={typeof, new, true, false, catch, function, return, null, catch, switch, var, const, let, extends, if, in, while, do, else, case, break, async, await,of,
  expect, field, toBeTruthy, toHaveLengthCondition, toBeAlphabetical, not, toBeEqual, fill, submit_form, assert, toBeNumerical
  },
  keywordstyle=\color{darkblue}\bfseries,
  ndkeywords={class, export, boolean, throw, implements, import, this, setTimeout},
  ndkeywordstyle=\color{brickred}\bfseries,
  identifierstyle=\color{black},
  sensitive=false,
  comment=[l]{//},
  morecomment=[f][\color{diffadd}\bfseries]{+\ },
  morecomment=[s]{/*}{*/},
  morecomment=[f][\color{diffremove}\bfseries]{- },
  commentstyle=\color{violet}\ttfamily,
  stringstyle=\color{carrotorange}\ttfamily,
  morestring=[b]',
  morestring=[b]"
}

\lstdefinelanguage{JSX}{
  keywords={Container, Navbar, Image, Input, Button, List, Card},
  keywordstyle=\color{darkblue}\bfseries,
  ndkeywords={name, src, placeholder, value, ...},
  ndkeywordstyle=\color{brickred}\bfseries,
  identifierstyle=\color{black},
  sensitive=false,
  comment=[l]{//},
  morecomment=[f][\color{diffadd}\bfseries]{+\ },
  morecomment=[s]{/*}{*/},
  morecomment=[f][\color{diffremove}\bfseries]{- },
  commentstyle=\color{violet}\ttfamily,
  stringstyle=\color{carrotorange}\ttfamily,
  morestring=[b]',
  morestring=[b]"
}

\theoremstyle{definition}

\newcommand{\header}[1]{\par\smallskip\noindent\textbf{#1.}}
\newcommand{\reducespace}{\vspace{-4mm}}

\usepackage{amsmath}

\AtBeginDocument{%
  }

\newboolean{showcomments}
\setboolean{showcomments}{true}
\ifthenelse{\boolean{showcomments}}
{
	\definecolor{myyellow}{RGB}{255, 228, 26}
	\definecolor{myblue}{RGB}{50, 50, 220}
	\newcommand{\nb}[2]{
		{\sf
			\fcolorbox{myyellow}{yellow}{\scriptsize\textbf{#1}}%
			$\blacktriangleright$%
			{\color{myblue}\fontsize{7pt}{8pt}\selectfont\textbf{#2}}%
		}%
	}
}
{
	\newcommand{\nb}[2]{}
}

\newboolean{showmaybe}
\setboolean{showmaybe}{true}
\ifthenelse{\boolean{showmaybe}}
{
	\definecolor{myyellow}{RGB}{255, 228, 26}
	\definecolor{myred}{RGB}{184, 37, 95}
	\newcommand{\maybe}[1]{
		{\sf
			\fcolorbox{myyellow}{yellow}{\scriptsize\textbf{Maybe}}%
			$\blacktriangleright$%
			{\color{myred}\fontsize{7pt}{8pt}\selectfont\textbf{#1}}%
		}%
	}
}
{
	\newcommand{\maybe}[1]{}
}

\algnewcommand\algorithmicforeach{\textbf{foreach}}
\algdef{S}[FOR]{ForEach}[1]{\algorithmicforeach\ #1\ \algorithmicdo}
\newcolumntype{Y}{>{\centering\arraybackslash}X}



\author{
    \IEEEauthorblockN{Parsa Alian}
    \IEEEauthorblockA{
        \textit{University of British Columbia}\\
        Vancouver, Canada \\
        palian@ece.ubc.ca
    }
    \and
    \IEEEauthorblockN{Martin Tang}
    \IEEEauthorblockA{
        \textit{University of British Columbia}\\
        Vancouver, Canada \\
        mt8422@student.ubc.ca
    }
    \and
    \IEEEauthorblockN{Ali Mesbah}
    \IEEEauthorblockA{\textit{University of British Columbia}\\
        Vancouver, Canada \\
        amesbah@ece.ubc.ca
    }
}

\pgfplotsset{compat=1.18}
\begin{document}

\title{
\approach: Inferring Component Abstractions\\ for Automated End-to-End Testing
}

\newcommand{\appname}{Amazon's web app\xspace}
\newcommand{\approach}{\textsc{Visca}\xspace}
\newcommand{\autoee}{\textsc{AutoE2E}\xspace}
\newcommand{\eebench}{\textsc{E2EBench}\xspace}
\newcommand{\vizmod}{\textsc{VizMod}\xspace}
\newcommand{\vips}{\textsc{VIPS}\xspace}
\newcommand{\gemini}{\textsc{Gemini 2.5 Flash}\xspace}

\newcommand{\code}[1]{{\small\ttfamily\texttt{#1}}}

\newcommand{\segmentationSubjects}{45\xspace}
\newcommand{\segmentationPrec}{83.4\%\xspace}
\newcommand{\segmentationRec}{23.3\%\xspace}
\newcommand{\segmentationF}{27.2\%\xspace}
\newcommand{\vipsHighPrec}{65.7\%\xspace}
\newcommand{\vipsHighRec}{60.8\%\xspace}
\newcommand{\vipsHighF}{48.5\%\xspace}
\newcommand{\countRatio}{3.5}
\newcommand{\vipsLowRatio}{0.4}
\newcommand{\vipsHighRatio}{8.2}
\newcommand{\segmentationImprov}{26.9\%\xspace}

\newcommand{\abstractionSubjects}{26\xspace}
\newcommand{\totalComponents}{386\xspace}
\newcommand{\structureMatchAverage}{91.4\%\xspace}
\newcommand{\structurePartialAverage}{5.7\%\xspace}
\newcommand{\structureNoAverage}{2.9\%\xspace}
\newcommand{\contentMatchAverage}{91.5\%\xspace}
\newcommand{\contentPartialAverage}{7.1\%\xspace}
\newcommand{\contentNoAverage}{1.4\%\xspace}
\newcommand{\completeness}{98.2\%\xspace}

\newcommand{\featureCoverage}{92\%\xspace}
\newcommand{\totalFeatureCoverage}{83\%\xspace}
\newcommand{\featureImprov}{16\%\xspace}

\newcommand{\autoeeCoverage}{79\%\xspace}

\maketitle

\thispagestyle{plain}
\pagestyle{plain}

\begin{abstract}
Providing optimal contextual input presents a significant challenge for automated end-to-end (E2E) test generation using large language models (LLMs), a limitation that current approaches inadequately address. This paper introduces Visual-Semantic Component Abstractor (\approach), a novel method that transforms webpages into a hierarchical, semantically rich component abstraction. \approach starts by partitioning webpages into candidate segments utilizing a novel heuristic-based segmentation method. These candidate segments subsequently undergo classification and contextual information extraction via multimodal LLM-driven analysis, facilitating their abstraction into a predefined vocabulary of user interface (UI) components. This component-centric abstraction offers a more effective contextual basis than prior approaches, enabling more accurate feature inference and robust E2E test case generation. Our evaluations demonstrate that the test cases generated by \approach achieve an average feature coverage of \featureCoverage, exceeding the performance of the state-of-the-art LLM-based E2E test generation method by \featureImprov.
\end{abstract}
\section{Introduction}
\label{sec:introduction}
Before deploying applications, developers need to thoroughly test them to ensure the quality and integrity of the features. End-to-end (E2E) testing, in particular, focuses on ensuring that all components of an application work together correctly and the features designed for the end-user perform smoothly.

While manual E2E test creation by human testers can yield high-quality, semantically rich tests, this approach is expensive, time-consuming, and scales poorly with application complexity, necessitating automated solutions. Traditional automated E2E testing techniques, often relying on application modeling~\cite{mesbah2012crawling, biagiola2019diversity, web-fragmentsRahul, web-matteo-icst20} or Reinforcement Learning (RL)~\cite{web-chang2023reinforcement}, aimed to reduce manual effort. However, the generated tests lack the quality and semantic coherence of human-written ones, often manifesting as random sequences of actions due to their reliance on criteria that do not inherently capture user-centric goals. 




Recent advances in Large Language Models (LLMs) have significantly expanded the potential for automating software testing across a variety of domains, including unit, API, and accessibility testing~\cite{nan2025test, cheng2025rug, yin2024you, kim2024multi, huq2024automated}. Building on this momentum, LLMs are also explored for E2E test generation. An emerging direction in this space involves leveraging LLMs to generate test cases that cover application features~\cite{alian2025feature}.
Feature inference in current techniques, such as \autoee~\cite{alian2025feature}, involves extracting HTML snippets of user actions and supplying them to an LLM to identify the underlying features. While intuitive, this strategy is hindered by insufficient contextual information. The HTML representation of an interactive element often lacks the surrounding semantics needed for the model to infer its role within the application. As a result, the absence of meaningful context can lead to inaccurate feature interpretations and limit the effectiveness of the generated E2E tests.

To address the challenge of finding optimal context for LLM-based feature inference in E2E testing, we introduce \textbf{Vi}sual-\textbf{S}emantic \textbf{C}omponent \textbf{A}bstractor (\approach). Our method transforms a given webpage into an abstract structured \textit{component abstraction}, which is analogous to the modular component models widely adopted in modern UI development frameworks such as React~\cite{react}, VueJS~\cite{vuejs}, and Svelte~\cite{svelte}. By encapsulating logically related structures, content, and behaviors, this higher-level component-centric abstraction captures the context of UI actions much more effectively than isolated raw HTML elements. Our insight is that this contextual abstraction, in turn, facilitates more accurate feature inference and the generation of more robust and meaningful E2E test cases.

The contributions of this work are as follows:

\begin{itemize}[leftmargin=*]
    \item Defining a standard vocabulary for describing the webpages as an abstraction.

    \item A heuristic-based segmentation technique, resulting in segments that are suitable for transformation to component abstractions.

    \item A novel method for transforming an unstructured webpage Document Object Model (DOM) into a hierarchical, semantically enriched component abstraction, identifying logical UI units and their semantic context.

    \item A feature inference technique that leverages the rich contextual information embedded within the component abstraction to more accurately identify the application features and generate test cases for them.
\end{itemize}

Our evaluations demonstrate the multi-faceted effectiveness of \approach. The core component abstraction module achieves a 91.5\% accuracy in preserving the structure and content of original applications. Within this, our novel heuristic-based segmentation algorithm attains a precision of \segmentationPrec, outperforming the state-of-the-art \vips algorithm by \segmentationImprov in this regard. Ultimately, for end-to-end test generation, \approach achieves an average feature coverage of \featureCoverage, surpassing the leading feature-driven E2E testing baseline by a \featureImprov improvement. These results collectively highlight the significant benefits of using detailed component abstractions to enhance contextual understanding for automated LLM-driven E2E test generation.

\section{Motivation}
\label{sec:motivation}

\begin{figure}[htbp]
    \centering
    \includegraphics[width=0.47\textwidth]{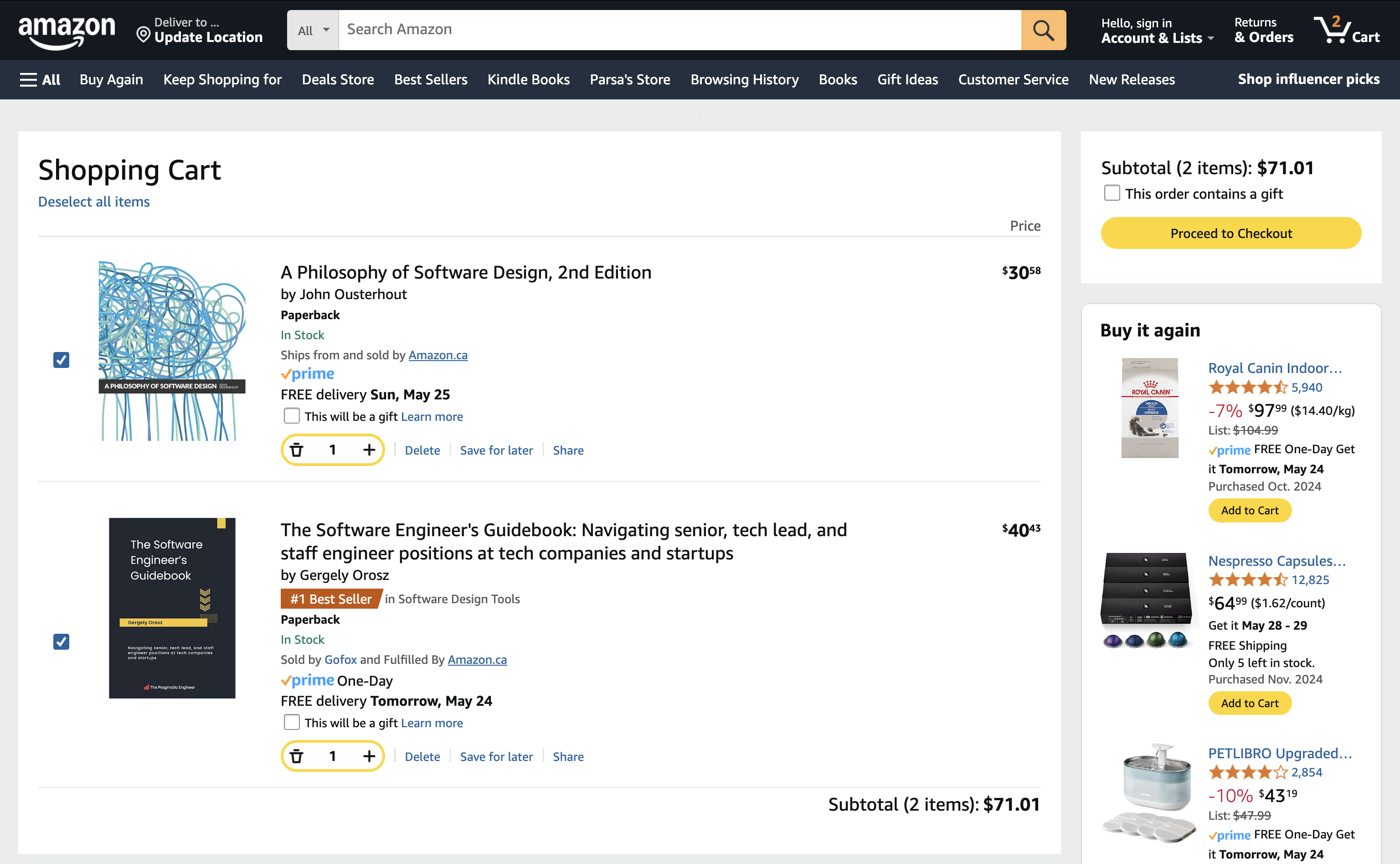}
    \caption{Amazon's shopping cart page}
    \label{fig:amazon-cart}
    \reducespace
\end{figure}

A critical factor in the success of LLM-based systems is the \textit{context} provided to the model; its nature and scope significantly influence whether an LLM performs a task effectively. For instance, AutoE2E utilizes the HTML of individual actions as its input context for feature inference. However, relying on such highly localized context presents considerable challenges. Consider Amazon's shopping cart page, depicted in \autoref{fig:amazon-cart}. On this page, users can ``delete a product from the cart." The simplified HTML for the ``Delete" action is shown in \autoref{lst:delete-action}.

\begin{lstlisting}[language=HTML, caption={Delete action's HTML from Amazon's cart page}, label={lst:delete-action}, basicstyle=\ttfamily\scriptsize, frame=tb, breaklines=true]
<span class="size-small action-delete">
  <input value="Delete" class="color-link"/>
</span>
\end{lstlisting}

Based on this action's HTML alone, the precise target of the ``Delete" is ambiguous. This snippet lacks the surrounding information necessary to determine its purpose. Furthermore, since multiple items are in the cart, their respective ``Delete" buttons share identical or
similar HTML. This makes it challenging for an LLM to distinguish which action corresponds to which item or to infer distinct features. In such cases, the provided context is \emph{too narrow}, potentially leading to inaccurate or incomplete feature inference.

To mitigate narrow context, another solution can be to provide the LLM with the entire page's HTML. While this theoretically grants access to all information needed to understand element relationships and infer features, it introduces new problems. Webpage HTML can span tens of thousands of lines of code; for example, the Amazon cart page (\autoref{fig:amazon-cart}) alone comprises roughly 36,000 lines of HTML, translating to approximately 348,000 LLM tokens. Feeding such extensive data to LLMs risks overlooking crucial information, increases the likelihood of model hallucination, and faces context window limitations. In this scenario, the context becomes \emph{too broad}, potentially obscuring relevant signals within excessive noise.

The challenge of selecting an optimal context is compounded by HTML's inherent design, primarily for presentation, not for semantic understanding. While the DOM provides a hierarchical structure, it offers no explicit cues for the ideal contextual boundary of any given interactive element. This ambiguity in defining the contextual scope makes it challenging to isolate a DOM segment that is both semantically complete for the task and efficiently processable.

To solve the problem of finding optimal context, one might consider segmentation to partition a webpage into distinct, visually and structurally coherent regions (e.g., headers, item lists, forms). Such segments could potentially offer more appropriately scoped, coherent contextual units for an LLM to analyze UI actions and infer features.
%
%
However, despite their potential, existing segmentation algorithms~\cite{cai2003vips, vineel2009web, kang2010repetition, kolcz2007site, meier2017fully, cormer2017towards, chen1906mmdetection, bajammal2021page} often fall short of providing optimal context for LLM-based feature inference. A primary limitation is their reliance on incomplete contextual analysis, as many such algorithms prioritize a single modality—such as DOM structure, textual content, or visual layout—each offering only a partial view when used in isolation. More critically, even when a collection of segments is generated, there is typically no inherent guidance on selecting which segments provide the appropriate contextual scope for a given UI element or feature. Individual segments can still be too broad, combining several distinct logical sections, or too narrow, fracturing a single coherent functional unit or providing only a partial view of an action's necessary context.

\begin{figure}[htbp]
    \centering
    \includegraphics[width=0.47\textwidth]{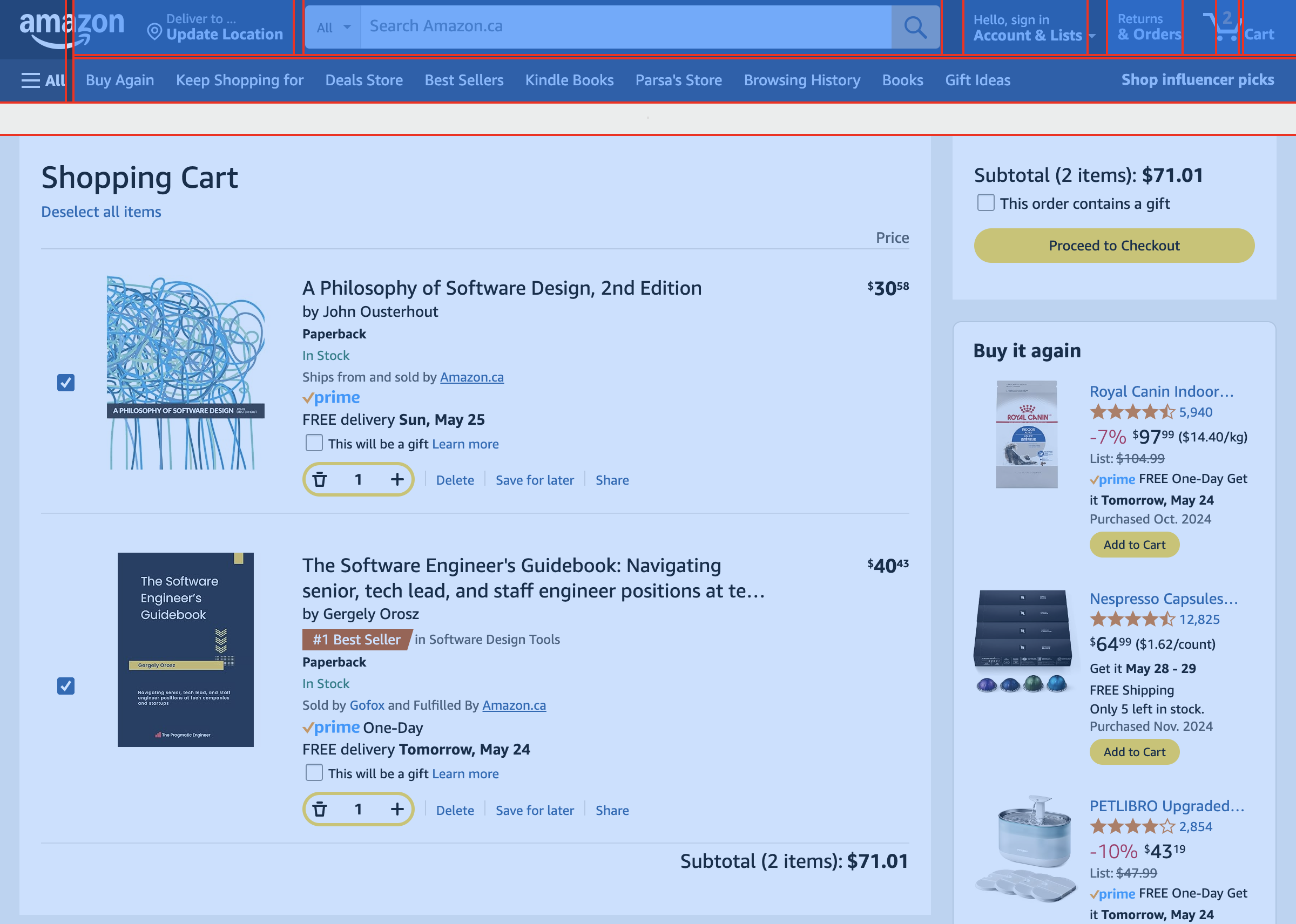}
    \caption{Amazon's cart page segmented by VIPS~\cite{cai2003vips}}
    \label{fig:amazon-cart-segmented}
    \reducespace
\end{figure}

\begin{figure*}[t]
    \centering
    \includegraphics[width=\textwidth]{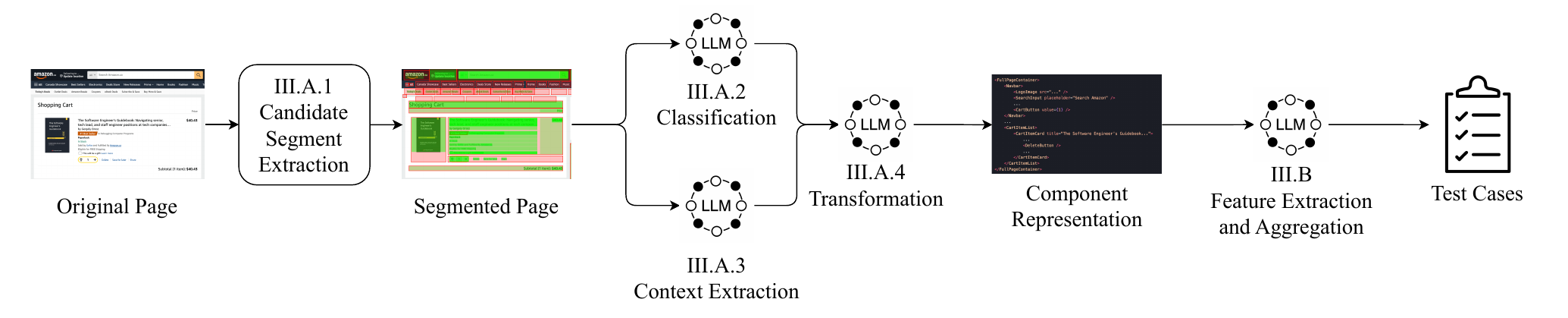}
    \caption{Overview of \approach} 
    \label{fig:workflow}
    \reducespace
\end{figure*}

For instance, \autoref{fig:amazon-cart-segmented} shows Amazon's cart page segmented by the VIPS algorithm~\cite{cai2003vips}, the state-of-the-art segmentation technique~\cite{kiesel2021empirical}. As depicted in the figure, the segment with the ``Delete" button (\autoref{lst:delete-action}) is excessively broad: it also includes the entire shopping cart, the checkout, and a ``Buy it again" section. This example demonstrates how a segmentation algorithm can produce an overly broad contextual unit.
Consequently, relying on segmentation outputs without a deeper contextual understanding of each segment is unreliable for providing the optimal LLM context.

We posit that an effective and balanced level of granularity for context can be found at the level of \emph{UI Components}.
Web application front-ends are built using HTML, CSS, and JavaScript. However, traditional development relying solely on these technologies often struggled with modularity, as HTML lacks inherent mechanisms for encapsulating the view and behavior of distinct application elements. Component-based frameworks (e.g., React, VueJS, Svelte) have emerged to address this, enabling developers to structure applications as collections of reusable, self-contained modules. Each component encapsulates a specific piece of functionality—like a cart item, a search bar, or a navigation menu—along with its associated UI elements, behavior, and styling.

UI components, by their very design, offer a more promising foundation for defining contextual boundaries than algorithmically derived segments. Developers create them as logically coherent and self-contained units. This means a component inherently groups related information and interactive elements that work together to fulfill a specific purpose. For example, an individual ``cart item" on the Amazon shopping cart page (\autoref{fig:amazon-cart}) would likely be implemented as a component. This component would naturally contain the product title, price, quantity controls, and the ``Delete" action, all related to that single entry. Such a component provides a context that is neither too narrow (it has all specifics for its actions) nor too broad (it is focused on one item), inherently forming a semantically meaningful boundary ideal for LLM comprehension.

\section{Approach}
\label{sec:approach}
To overcome the limitations of insufficient context for robust feature inference, we introduce \approach. Our method transforms webpages into a semantically rich, component-based abstraction through a multi-stage process, and then leverages this abstraction for feature inference and test case generation. The \approach workflow, illustrated in \autoref{fig:workflow}, begins by processing nodes from the DOM tree, applying heuristics to generate an initial hierarchy of candidate segments. These candidates are subsequently classified while simultaneously extracting contextual descriptions using multimodal analysis. The classified segments are then transformed into a component abstraction. Utilizing the structure and descriptions provided by these components, we infer application features and automatically generate corresponding E2E test cases.

\subsection{Transforming to Component Abstraction}
\label{sec:segmentation}

To facilitate modular development over raw HTML, CSS, and JavaScript development, component-based frameworks enabled the encapsulation of logic, structure, and style into discrete \textit{UI Components}. Building upon this paradigm, popular design libraries, such as those detailed in \autoref{tab:component-libraries}, provide collections of widely-used \textit{Common Templates}. These templates offer predefined, reusable implementations of UI components that adhere to established design patterns, facilitating more efficient and consistent application development by reducing the need to build every interface element from scratch.

\begin{table}[t]
{\small
    \centering
    \caption{Component Libraries}
    \label{tab:component-libraries}
    \begin{tabular}{llrr}
        \toprule
        \textbf{Library Name} & \textbf{Component Count} & \textbf{GitHub Stars} \\
        \midrule

        \rowcolor{lightgray}
        Bootstrap~\cite{bootstrap} & 25 & 172K \\

        Material UI~\cite{mui} & 50 & 95K \\

        \rowcolor{lightgray}
        Ant Design~\cite{antd} & 70 & 94K \\

        Shadcn~\cite{shadcn} & 49 & 85K \\

        \rowcolor{lightgray}
        Chakra UI~\cite{chakra} & 95 & 39K \\
        \bottomrule
    \end{tabular}
}
\reducespace
\end{table}


Given that common templates are designed for broad applicability across diverse websites, we posit that they can also serve as a descriptive vocabulary to \textit{express} the structure and function of virtually any webpage, regardless of its original implementation technology or framework usage. For instance, consider Amazon's shopping cart page, depicted in \autoref{fig:amazon-cart}. This page was not built with any of the specific libraries listed in \autoref{tab:component-libraries}, and its underlying HTML presents a deeply nested structure composed of semantically generic elements. Nevertheless, it is possible to conceptualize an abstraction of this page using common templates analogous to those found in such libraries:

\begin{lstlisting}[language=JSX, caption={Abstraction of Amazon's shopping cart page}, label={lst:amazon-abstraction}, basicstyle=\ttfamily\scriptsize, frame=tb, breaklines=true]
<Container name="Shopping Cart Page">
  <Navbar>
    <Image name="Logo" src="..." />
    <Input name="Search" placeholder="Search Amazon" />
    ...
    <Button name="Cart" value="2" />
    ...
  </Navbar>
  ...
  <List name="Shopping Cart Item List">
    <Card name="Shopping Cart Item" title="A Philosophy of Software Design...">
      ...
      <Button name="Delete Cart Item" />
      ...
    </Card>
    <Card name="Shopping Cart Item" title="The Software Engineer's Guidebook...">
      ...
      <Button name="Delete Cart Item" />
      ...
    </Card>
  </List>
  ...
</Container>
\end{lstlisting}

The abstracted form in \autoref{lst:amazon-abstraction} is designed such that if rendered using UI libraries (\autoref{tab:component-libraries}), the resulting webpage would be visually and structurally congruent with the original view in \autoref{fig:amazon-cart}. This abstraction, unlike the page's raw HTML, offers significant advantages for contextual understanding. It directly addresses the limitations outlined in \autoref{sec:motivation} by semantically grouping related content and behavior, thereby aiding LLMs in feature inference. For instance, each ``cart item" is represented as a distinct \code{<Card name="Cart Item">} in this abstraction. This structured representation delineates boundaries for actions within each conceptual unit.

Moreover, each component instance inherently carries semantic context based on its type. For example, the \code{Navbar} presented in \autoref{lst:amazon-abstraction} immediately implies that its children are navigation-related functionality and content. Such an understanding provides richer information than analyzing raw HTML elements alone.

Therefore, \approach first generates a \textit{component abstraction} (similar to \autoref{lst:amazon-abstraction}) using a curated set of \textit{common templates}. These templates were systematically derived by analyzing widely-used UI design libraries (\autoref{tab:component-libraries}), focusing on elements with clear semantic and functional roles while consolidating synonyms and generalizing overly specialized items to ensure broad applicability. This curation resulted in a standardized vocabulary of 50 common templates and their names.
Examples of these common templates include \code{Card}, \code{Table}, \code{Navbar}, or \code{Tab} components.

The resulting abstraction serves as semantically rich, context-aware segments, providing a more meaningful foundation for feature inference compared to raw DOM analysis or traditional segmentation methods. The following sections detail the multi-stage process \approach employed to achieve this abstraction.

\subsubsection{Candidate Segment Extraction}
\label{sec:candidate-segments}

The initial DOM tree of a webpage is often large and computationally intensive to analyze. Therefore, \approach initially focuses on simplifying this into a more manageable set of \textit{Candidate Segments}. This begins by identifying all visible DOM nodes and reconstructing their hierarchical relationships. Subsequently, a recursive, visual process prunes structurally redundant nodes, added for styling or layout purposes, that offer no semantic value distinct from their children. This process iteratively examines parent nodes with exactly one child: if the parent's visual rendering is identical to, or a close derivative of, its child's (e.g., differing only by padding or minor positional shifts), the child node is considered redundant and is removed. The children of any pruned node are then re-parented to their grandparent.

While the pruning step reduces the complexity of the initial DOM hierarchy, the number of remaining nodes can still be substantial for complex webpages. Processing a large number of fine-grained nodes in subsequent LLM-based stages might be computationally expensive or inefficient. To address this potential scalability challenge, \approach includes a heuristic-based segmentation step. This step aims to identify a set of candidate segments by selecting nodes that likely represent component instances based on two key principles: appropriate \textit{size} and structural \textit{similarity} to siblings, reflecting common patterns in component-based design.

First, component instances typically encapsulate a meaningful unit of content or functionality, suggesting they should not be granular (e.g., single text nodes) nor large (e.g., entire page sections). The idea of balancing segment size has been explored in techniques like \vizmod~\cite{bajammal2018generating} for wireframe to component transformation, where the approach considers a node's subtree size in its heuristics. Our approach similarly favors nodes representing reasonably sized subtrees.


Second, developers leverage component templates for reusability. This results in multiple instances of the same component template appearing adjacent or repeated within a section. An example of this can be seen in the repeated use of \textit{Cart Item} in Amazon's shopping cart page (\autoref{fig:amazon-cart}). Therefore, a node representing a component instance is likely to be structurally similar to its siblings if they originate from the same template.

To quantitatively capture both size and sibling similarity, we define the following potential function $\Psi_n$ for each node $n$ in the pruned hierarchy from the previous step:
\begin{equation}
\label{eq:potential_function}
\Psi_n \;=\; \log\!\Biggl(
  \frac{\mathrm{size}(n)}
       {1 + \displaystyle\sum_{i}\mathrm{dist}\bigl(n,\mathrm{sib}_i(n)\bigr)}
\Biggr)
\end{equation}

In this function, $\displaystyle \mathrm{size}(n)$ measures the size of the subtree rooted at node $n$, and 
$\displaystyle \Sigma\mathrm{dist}\bigl(n,\mathrm{sib}_i(n)\bigr)$ denotes the total structural dissimilarity between node $n$ and its sibling nodes $\mathrm{sib}_i(n)$.
This is calculated using an appropriate tree comparison metric, such as Tree Edit Distance~\cite{zhang1989simple}, summed over all siblings. The function $\Psi_n$ assigns higher potential to nodes that are larger and similar to their siblings. To select the candidate segments based on this potential function, we employ Algorithm~\ref{alg:mark_candidates}.

\begin{algorithm}
\caption{Candidate Segment Extraction Algorithm}
\label{alg:mark_candidates}
{\scriptsize
\begin{algorithmic}[1]
\Function{MarkCandidateSegments}{node}
    \State $\textit{nodePotential} \gets \text{CalculatePotential}(\textit{node})$ \Comment{Calculate $\Psi_{\textit{node}}$ using Eq.~\eqref{eq:potential_function}}
    \State $\textit{childrenCombinedPotential} \gets 0$
    \ForAll{child \textbf{in} node.Children}
        \State $\text{MarkCandidateSegments}(\textit{child})$
        \State $\textit{childrenCombinedPotential} \text{+=} \textit{child}\text{.assignedPotential}$
    \EndFor 
    \If{$\textit{nodePotential} \ge \textit{childrenCombinedPotential}$}
        \State $\textit{node}.\text{isCandidate} \gets \textbf{true}$
        \State $\textit{node}.\text{assignedPotential} \gets \textit{nodePotential}$
        \ForAll{descendant \textbf{in} node.Descendants}
            \State descendant.isCandidate = false
        \EndFor
    \Else
        \State $\textit{node}.\text{isCandidate} \gets \textbf{false}$
        \State $\textit{node}.\text{assignedPotential} \gets \textit{childrenCombinedPotential}$
    \EndIf
    \State \textbf{return} $\textit{node}$
\EndFunction
\end{algorithmic}
}
\end{algorithm}

The \code{MarkCandidateSegments} function first calculates the potential ($\Psi_n$) for the current \code{node} (line 2). It then recursively calls itself on all children and aggregates the \code{assignedPotential} evaluated by these calls (lines 3-7). This \code{assignedPotential} represents the maximum potential achievable within each child's subtree (either from the child itself or its descendants). The algorithm compares the node's own potential (\code{nodePotential}) with the
sum passed up from its children (\code{childrenCombinedPotential}, line 8). If the node's potential is higher or equal, it indicates that this node is a better candidate than any combination of candidates within its children. Therefore, the node is marked (\code{isCandidate = true}), its own potential is assigned, and any candidates previously marked within its subtree are unmarked  (lines 12). If the children's combined potential is higher, the node itself is not marked, and the combined potential from the children is assigned to the node to be passed up the recursion (lines 15-16). After the traversal completes on the root, the set of nodes with \code{isCandidate = true} represents the heuristically selected candidate segments.

\subsubsection{Candidate Segment Classification}
\label{sec:candidate-classification}
Following the candidate segment extraction, \approach classifies each candidate segment to determine its semantic role. This classification is crucial for identifying which segments represent UI component instances suitable for transformation into the target component abstraction and understanding the overall page structure. We categorize candidate segments into three distinct types: Containers, Lists, and Components.

\header{Containers}
The primary function of a container is organizational; it groups various subsegments based on spatial layout or structural hierarchy without enforcing strong semantic cohesion among the contained elements. An example of a Container is the \vips segment in \autoref{fig:amazon-cart-segmented}, containing the shopping cart, checkout section, and Amazon's reward program.

\header{Lists}
The second candidate segment category is the \textit{List}. While Lists, like Containers, serve a grouping function, they are distinguished by stronger internal semantic cohesion: the subsegments organized within a List represent multiple instances of the same conceptual entity or fulfill the same semantic role. An example of lists includes the ``cart item"  list shown in \autoref{fig:amazon-cart}.

\header{Components}
The final category, and the primary target for our subsequent transformation, is the \textit{Component}. Components are instances of common templates, representing a coherent unit of information and functionality. An example of a Component includes a single ``cart item" (\autoref{fig:amazon-cart}).

\begin{figure}
    \centering
    \includegraphics[trim={0.5cm 0.5cm 0.5cm 0.5cm},clip,width=0.35\textwidth]{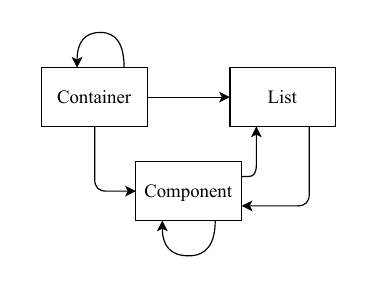}
    \caption{Segment types' composition relations} 
    \label{fig:segment-relations}
    \reducespace
\end{figure}

The defined segment categories exhibit a hierarchical containment relationship, illustrated in \autoref{fig:segment-relations}. Containers function as general structural groupings and may contain any segment type. Lists, enforcing semantic consistency among their children, and contain only multiple instances of a single type of Component. Finally, Components support recursive composition, allowing them to encapsulate nested Lists or other smaller Components.

To assign a category to each candidate segment produced by the preceding steps (Section \ref{sec:candidate-segments}), \approach utilizes prompting of a multimodal LLM. This step determines the semantic role of each segment and identifies those suitable for transformation into our target component abstraction. For each candidate segment, the multimodal LLM is provided with the definitions and hierarchical relationships of the three segment categories, and the visual rendering of the candidate segment. We employ Chain-of-Thought (CoT) prompting to guide the LLM's reasoning process. The prompt encourages the LLM to analyze the segment's visual characteristics, considering factors like internal homogeneity of the content and structure to determine the most appropriate category.

We use the recursive procedure detailed in Algorithm~\ref{alg:tree_classification} to apply this classification efficiently across the entire candidate segment hierarchy.

\begin{algorithm}
\caption{Segment Tree Classification}
\label{alg:tree_classification}
{\scriptsize
\begin{algorithmic}[1]
\Function{ClassifyTree}{node}
    \State \textit{node}.\text{class} $\gets$ \Call{MultimodalLLMClassify}{\textit{node}}
    \If{$\textit{node}.\text{class} = \text{``Container"}$}
        \ForAll{child \textbf{in} node.Children}
            \State \Call{ClassifyTree}{\textit{child}}
        \EndFor
    \ElsIf{$\textit{node}.\text{class} = \text{``List"}$}
        \ForAll{child \textbf{in} node.Children}
            \State $\textit{child}.\text{class} \gets \text{``Component''}$
        \EndFor
    \EndIf
    \Comment{Otherwise node.class is Component: Base case, recursion stops down this path}
    \State \textbf{return} \textit{node}
\EndFunction
\end{algorithmic}
}
\end{algorithm}

The algorithm begins by invoking the multimodal LLM process (\code{MultimodalLLMClassify}) to determine the class for the current \code{node} (line 2). The subsequent steps depend on the assigned class. If a node is classified as a Container (line 3), its children can be of any category. Therefore, the algorithm must recursively call itself on each child to determine their classes (lines 4-6). If a node is classified as a List (line 7), its children are, by definition of our List category, expected to be multiple instances of a Component. The algorithm thus directly assigns the \code{Component} class to these children (lines 8-10). Finally, if a node is classified as a Component, it represents a structural leaf in terms of this classification process. No further recursive calls are made down this branch, as the segment is now identified as a candidate for the final transformation step.


\subsubsection{Candidate Segment Context Extraction}
\label{sec:context-extraction}
A primary motivation for component abstraction is acquiring richer context for feature inference and test generation (Section \ref{sec:motivation}). In practice, developers often assign descriptive names to UI components (e.g., \code{<CartItemCard>} or \code{<SearchResultCard>} rather than a generic \code{<Card>}), improving the understandability of the application structure. To achieve a similar level of semantic insight, \approach extracts a title and a natural language contextual description for each candidate segment, performing this task simultaneously with the classification process described in Section \ref{sec:candidate-classification}.
We utilize the following prompt structure to extract the contextual description for the candidate segment:

\begin{enumerate}[leftmargin=*]
    \item \textbf{Page Context:} The overall page description is generated based on the full-page visual rendering, providing a contextual understanding of the application under test.
    
    \item \textbf{Visual Rendering:} The candidate segment's visual rendering serving as the primary source for its content and layout.
    
    \item \textbf{Ancestor Context:} The contextual description(s) previously generated for the current segment's ancestors. Knowing that a segment containing the product information is nested within ``\textit{Shopping Cart}" (\autoref{fig:amazon-cart}) helps clarify its role compared with when the same information is in other possible contexts.
\end{enumerate}

To determine these contextual descriptions across the hierarchy, we employ a top-down recursive approach. By combining page-level, ancestor, and local visual context, \approach aims to generate descriptions that accurately reflect the specific role and purpose of each segment. An example of the final generated contextual description for the ``cart item" from \autoref{fig:amazon-cart} can be seen in \autoref{lst:sample-context}.

\begin{lstlisting}[language=JavaScript, caption={Extracted context for cart item}, label={lst:sample-context}]
{
    "title": "Shopping Cart Item",
    
    "context": "This segment displays the detailed information and interactive options for a product in a shopping cart or order summary."
}
\end{lstlisting}

These generated contextual descriptions are subsequently associated with each respective segment in its final abstraction. For example, in the abstracted structure illustrated in \autoref{lst:amazon-abstraction}, this concept is exemplified by each node featuring a \code{name} attribute that encapsulates its derived contextual understanding.

\subsubsection{Candidate Segment Transformation}
\label{sec:transformation}
The final step in the segmentation phase of \approach is to generate the hierarchical component abstraction. This structured abstraction serves as the enriched input for the feature inference and test generation stage. The specific transformation strategy employed depends directly on the category assigned to each segment during classification: 
%
%
%
For \emph{Containers}, we recursively generate the abstraction for each child segment it contains and then assemble these child abstractions according to the parent’s structure. When handling \emph{Lists}, we avoid redundancy by selecting a single child segment as a representative sample; the component abstraction is then generated solely for this sample, which defines the structure and type for all similar items in the original list. Lastly, segments that are directly classified as \emph{Components} serve as the base level for our transformation and are generated through a multimodal LLM prompting process.

The core transformation in \approach occurs for segments classified as \textit{Components}. We employ a multimodal LLM prompt to map each such segment onto the most appropriate common templates from our predefined vocabulary.
To guide the LLM towards an accurate transformation, the prompt integrates several key pieces of information for each target \textit{Component} segment:

\begin{enumerate}[leftmargin=*]
    \item \textbf{Segment's Visual Rendering:} The visual rendering provides the primary evidence of the component's appearance, layout, and visible sub-elements.
    
    \item \textbf{Segment HTML Code:} The corresponding HTML source is included to allow the LLM to extract precise textual content, specific attributes (e.g., \code{href} for URLs, \code{src} for image paths), or structural nuances potentially missed by visual analysis alone.
    
    \item \textbf{Hierarchical Context:} The titles and the natural language descriptions generated for the segment's ancestors (Section \ref{sec:context-extraction}) are provided. This informs the LLM about the component's placement and likely role within the broader application structure, aiding disambiguation (e.g., differentiating visually similar cards used for different purposes).
    
    \item \textbf{Common Template Definitions:} The list of the 50 target common templates is provided to the LLM. This grounds the generation process, instructing the LLM to structure its output according to this predefined vocabulary.
\end{enumerate}

Once component abstractions have been generated for the relevant segments, \approach completes the page transformation via a bottom-up composition phase. This recursive assembly culminates at the hierarchy's root node; its final, fully composed abstraction constitutes the complete component-based model derived from the original web page, ready to be used in the subsequent feature inference and test generation phase. The result of this transformation is an abstracted version of the web page, such as the example illustrated in \autoref{lst:amazon-abstraction} for Amazon's shopping cart page (\autoref{fig:amazon-cart}).

\subsection{Test Generation}
\label{sec:test-generation}
The final phase of \approach utilizes the generated component abstraction to drive feature inference and E2E test case generation. For this, we adapt the probabilistic framework established by \autoee~\cite{alian2025feature}. As previously noted (Section \ref{sec:motivation}), the original \autoee framework processes the raw HTML of individual UI actions to infer features. Our primary modification is to replace this direct HTML input with our component abstraction inference. Consequently, instead of analyzing isolated action elements, the adapted framework now leverages the semantically richer components to infer application features and subsequently generate targeted E2E test cases. 

\subsection{Implementation}
\label{sec:implementation}
\approach is implemented in Python, leveraging \gemini as its core LLM. We selected \gemini primarily for its affordability when operating at larger scales, which facilitates the practical application of \approach in real-world test generation scenarios. To promote consistent and reproducible outcomes, \gemini was utilized with a temperature setting of 0. Furthermore, although \gemini natively supports reasoning capabilities, all operations within \approach that generate the results presented in this work were performed using its non-reasoning version. The features inferred for the components and the generated test cases are stored on \textsc{MongoDB Atlas}~\cite{mongodb}. The final test cases generated for the features utilize the Selenium~\cite{selenium} framework for testing and assertions.
\section{Evaluation}
\label{sec:evaluation}
To measure the effectiveness of our approach, we address the following research questions.

\begin{itemize}[leftmargin=*]
\item \textbf{RQ1}: How accurate is the component abstraction module in \approach? 

\item \textbf{RQ2}: How effective is the heuristic-based segmentation algorithm?

\item \textbf{RQ3}: How does \approach compare to the state-of-the-are E2E test generation methods?
\end{itemize}

\subsection{Component Abstraction Accuracy (RQ1)}
\label{sec:eval-abstraction}
To evaluate the accuracy of our component abstraction module, the core of \approach, we need to compare its generated component-based abstraction against a ground-truth abstraction that also utilizes our defined common templates. One practical method to achieve this is to obtain subject webpages that were originally constructed using UI components directly analogous to these common templates. For such pages, their inherent, library-defined component structure effectively serves as the ground-truth abstraction. \approach's generated abstraction for these pages can then be compared against this known structure to determine how accurately it captures the intended component-based design.

Following this methodology to construct our ground-truth dataset, we investigated the open-source UI libraries listed in \autoref{tab:component-libraries}. Our selection process involved examining their official documentation and publicly available showcase applications, as these directly exemplify pages built with well-defined UI components. Among the five UI libraries we initially considered (\autoref{tab:component-libraries}), three provided open-source demonstration pages suitable for our analysis: Material UI, Shadcn, and Bootstrap. Our selection process was as follows: we included all available showcase pages from both Bootstrap and Shadcn. For Material UI, given that its operational showcases primarily feature dashboard interfaces, we selected two distinct dashboard applications and incorporated all unique webpages within them. This methodology yielded \abstractionSubjects distinct webpages that serve as our evaluation subjects. 

Each subject is thus a webpage natively constructed using UI components from its respective library. Crucially for establishing the ground truth, the library-specific component types (e.g., markup explicitly defining a \code{Button} or a \code{Card}) are embedded within the subject's HTML code. This embedded type information, alongside the page's complete HTML structure and visual rendering, constitutes the ground truth against which the component abstractions generated by \approach are evaluated.

\header{Methodology}
For each subject webpage, we execute the component abstraction phase of \approach. The input to \approach consists of the page's raw HTML code and its visual rendering. Crucially, to assess \approach's ability to generalize, any explicit library-specific component identifiers (e.g., attributes like \code{class="card"}) present in the original subject HTML are omitted from the input. The output of this phase is a component abstraction for the page (similar to \autoref{lst:amazon-abstraction}).

The evaluation then focuses on the segments that \approach has identified and transformed into instances of our \textit{Component} category (as defined in Section \ref{sec:candidate-classification}). For each such generated Component instance, we compare its abstracted representation against the corresponding ground-truth component's original HTML snippet from the subject page. This comparison is performed across two primary dimensions: \textit{structural accuracy} and \textit{content preservation}.

Structural accuracy assesses how well the common template selected by \approach and its structure align with the actual type and structural characteristics of the ground-truth component instance (e.g., a ground-truth \code{Card} component is represented in the abstraction using \code{Card} and its structure). Content preservation verifies whether all essential textual content, interactive elements, and key functional attributes (e.g., hyperlink URLs, image sources, button labels) from the ground-truth component's HTML are accurately and completely captured within \approach's generated component abstraction. Each comparison for structural accuracy and, separately, for content preservation, yields one of the following three outcomes:

\begin{itemize}[leftmargin=*]
    \item \textbf{Full Match:} For \textit{structural accuracy}, this indicates the component abstraction's type and essential structure perfectly align with, or are an appropriate generalization of, the ground-truth component. For \textit{content preservation}, this signifies that all essential information (text, key attributes, interactive elements) from the source HTML is accurately and completely captured in the abstraction.

    \item \textbf{Partial Match:} For \textit{structural accuracy}, the component abstraction does not exactly match the ground-truth type or precise structure, but it represents a semantically plausible alternative or a reasonable generalization that could achieve a similar visual and functional outcome (e.g., using a \code{List} template for navigation options instead of a specific \code{Navbar} template, if the navigational purpose is maintained). For \textit{content preservation}, most essential content is present, but some minor attributes, secondary text values, or non-critical elements are missing or differ slightly in the abstraction, without fundamentally altering the core information or function of the component.

    \item \textbf{No Match:} If the criteria for neither a Full Match nor a Partial Match are met for the dimension being evaluated (either structure or content), the abstraction is considered incorrect in that specific aspect.
\end{itemize}

\begin{figure}[t]
    \centering
    \includegraphics[width=0.48\textwidth]{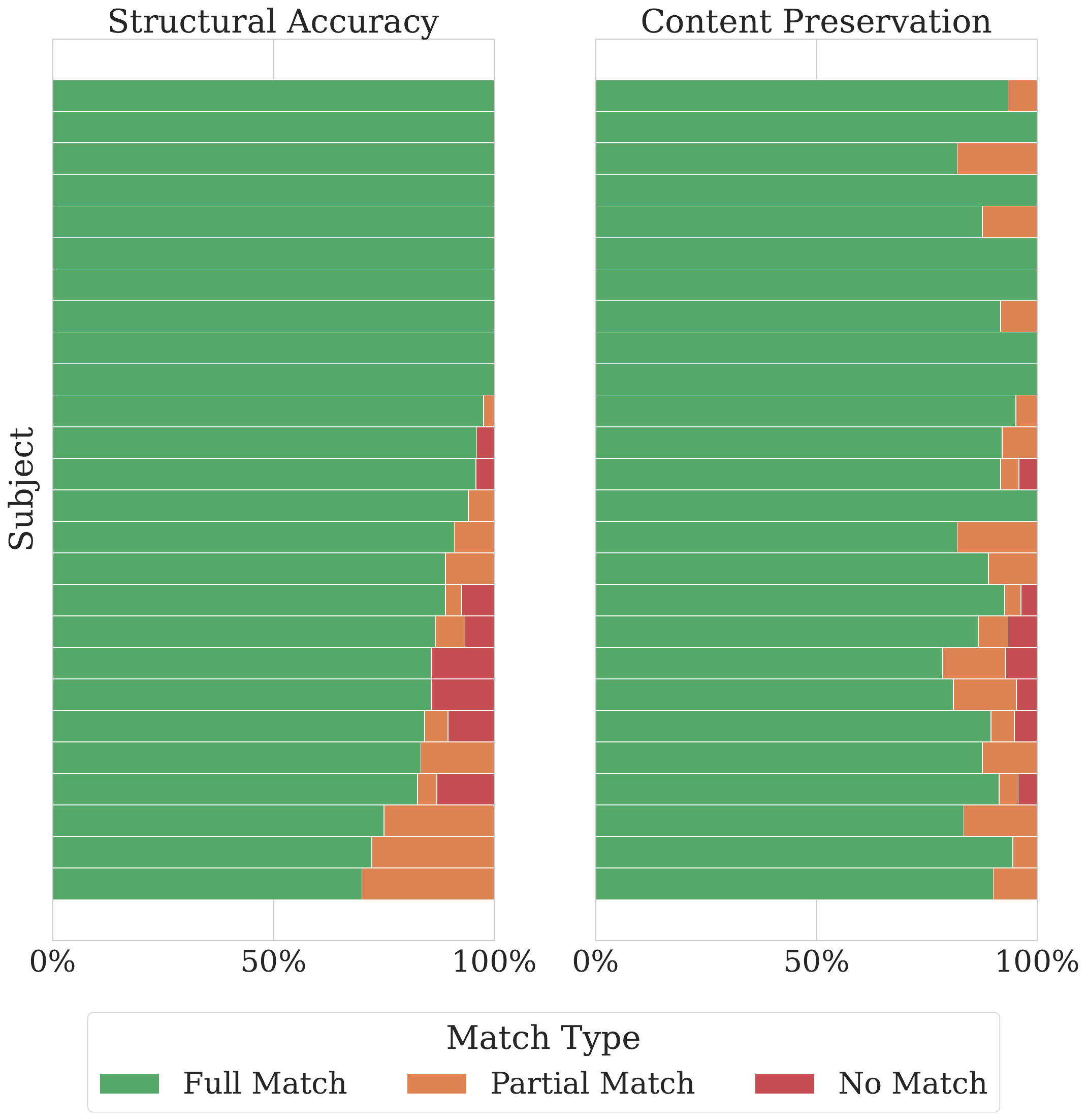}
    \caption{Abstraction Effectiveness for \approach} 
    \label{fig:abstraction}
    \reducespace
\end{figure}

To determine the match category (Full, Partial, or No Match) for both structural accuracy and content preservation, we first employ \gemini, configured in its reasoning mode. For each abstraction generated by \approach, \gemini is provided with the abstraction, its original HTML code, and a visual rendering of the original segment. The LLM is then prompted to assess these inputs against our defined criteria and assign the appropriate match category for structure and, separately, for content. Following the LLM's assessment, the assigned match category for each comparison is independently validated by multiple authors to ensure the accuracy and consistency of the final evaluation.

The results for structural accuracy and content preservation are presented in \autoref{fig:abstraction}. Each bar in this figure illustrates the percentage of generated abstract \textit{Component} in different match categories. In total, \totalComponents components have been generated by \approach across the \abstractionSubjects webpages. For structural accuracy, \structureMatchAverage of these components are a full match, \structurePartialAverage partial match, and \structureNoAverage no match. For content preservation, \approach achieves a \contentMatchAverage full match, \contentPartialAverage partial match, and \contentNoAverage no match. This means that in total, around 91.5\% of the components generated by \approach are perfect representations of the underlying UI component used in creating them.

To ensure that \approach accounts for all relevant parts of the page, we measure the \textit{completeness} of the abstraction. This metric calculates the proportion of the visual elements from the subject's ground-truth component structure that are not represented in our generated abstraction. A lower proportion of absent elements indicates a more comprehensive transformation. In total, \approach only missed seven components in its abstractions in the entire dataset, meaning that the completeness of \approach is evaluated at \completeness.

\subsection{Segmentation Performance (RQ2)}
\label{sec:eval-segmentation}
Our approach incorporates a novel heuristic-based segmentation method, detailed in Section \ref{sec:candidate-segments}. We conduct two primary evaluations for this segmentation technique. First, we assess its overall segmentation effectiveness by comparing it against a state-of-the-art baseline.
To establish a robust baseline for comparison, we referred to previous empirical studies of popular webpage segmentation algorithms~\cite{kiesel2021empirical}. This prior research identifies the \vips algorithm~\cite{cai2003vips} as a top-performing method overall. Consequently, we benchmark our segmentation heuristic's performance against \vips.

To evaluate the segmentation results, we utilize a dataset of human-annotated real-world websites~\cite{kreuzer2013quantitative}, which closely mirrors the complex DOM structures found on live websites. This dataset contains the top \segmentationSubjects most popular websites from Yahoo's Web Directory, where each page has been manually divided into semantic segments. As evaluation metrics, we employ $P_{B^3}$ (precision), $R_{B^3}$ (recall), and $F_{B^3}$ ($F_1$-score), which were previously introduced for assessing web page segmentation results~\cite{kiesel2020web}. These metrics assess segmentation by treating the segments as clusters and measuring their similarity based on shared pairs, a metric known as B-Cubed~\cite{amigo2009comparison}. 





\begin{figure}
    \centering
    \includegraphics[width=0.48\textwidth]{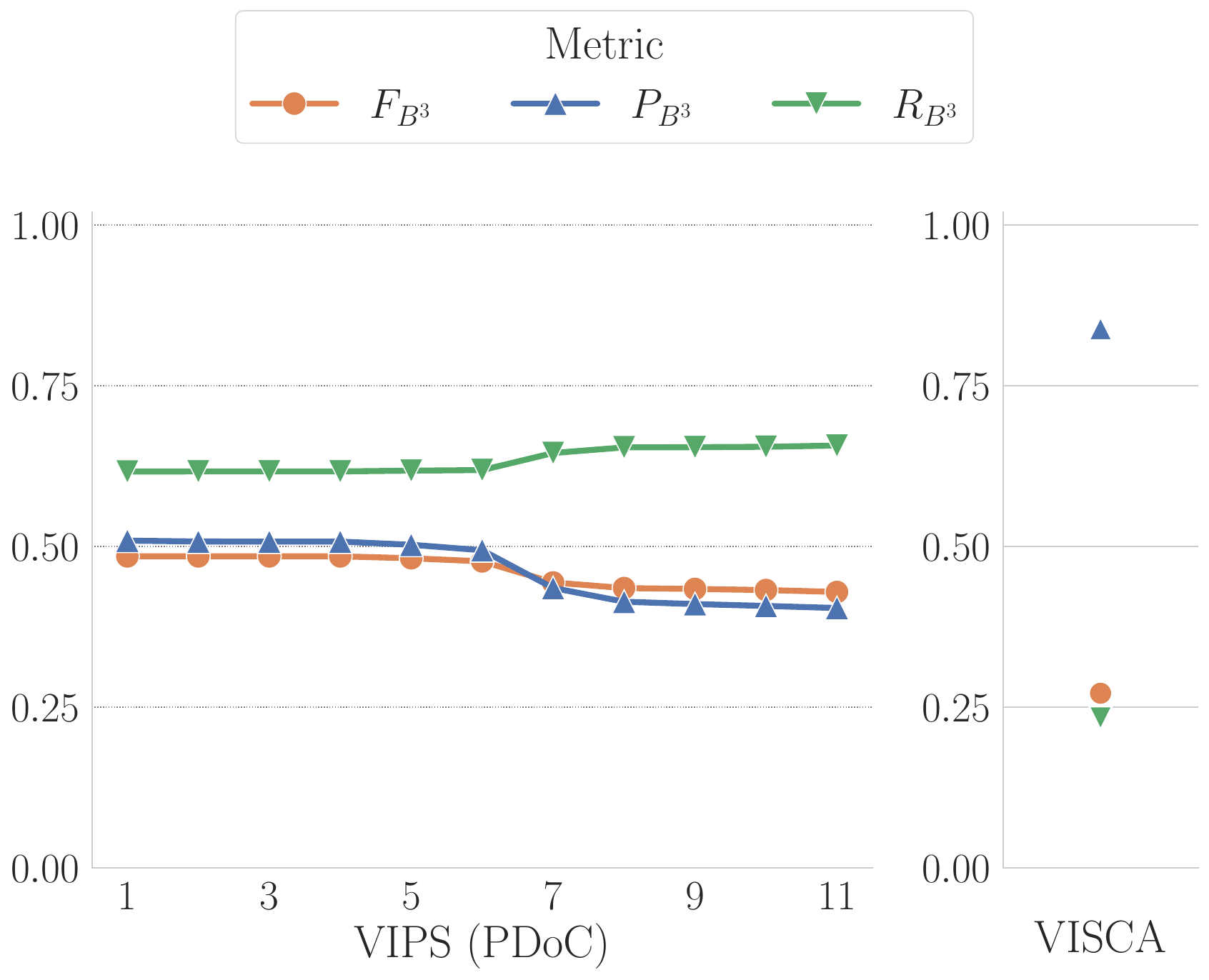}
    \caption{Comparison of B-Cubed scores for \vips and \approach}
    \label{fig:segmentation-results}
    \reducespace
\end{figure}

\autoref{fig:segmentation-results} displays the results of our segmentation evaluation, comparing \approach with the \vips baseline using the $P_{B^3}$ (precision), $R_{B^3}$ (recall), and $F_{B^3}$ ($F_1$-score) metrics. The \vips algorithm's performance is tuned via its Permitted Degree of Coherence (PDoC) parameter, which is a threshold determining whether the algorithm should divide a segment into smaller subsegments, influencing the size, number, and coherence of the segment's content.
We varied PDoC in its allowed range from 1 to 11, reporting its strongest performance. As illustrated, \approach achieves a $P_{B^3}$ of \segmentationPrec, an $R_{B^3}$ of \segmentationRec, and an $F_{B^3}$ of \segmentationF. For comparison, the optimal $P_{B^3}$, $R_{B^3}$, and $F_{B^3}$ scores achieved by \vips across all tested PDoC values are \vipsHighPrec, \vipsHighRec, and \vipsHighF, respectively. Our method's precision outperforms the best-performing \vips's precision by \segmentationImprov.

The primary objective of our segmentation heuristic (Section \ref{sec:candidate-segments}) is to generate candidate segments that align closely with our \textit{Component} category, as defined in Section \ref{sec:candidate-classification}. To assess how well we achieve this goal, we evaluate the proportion of segments produced by our heuristic that are subsequently identified as \textit{Components} during our multimodal LLM classification stage (Section \ref{sec:candidate-classification}). We compare this proportion against that achieved by the \vips algorithm. For \vips, we consider its output at three PDoC settings: 1 and 11 (the range extremes), and 6, which has been identified as a generally optimal parameter~\cite{kiesel2021empirical}. The results of this comparative analysis are presented in \autoref{tab:classification-stats}.

\begin{table}[h]
{\small
    \caption{Classification Statistics for \vips and \approach Segments}
    \label{tab:classification-stats}
    \centering
    \begin{tabular}{lrrr}
        \toprule
        \textbf{Method} & \textbf{Avg Segments} & \textbf{Component} & \textbf{Non-Component} \\
        \midrule

        \rowcolor{lightgray}

        $\vips_1$ & 10 & 43.0\% & 57.0\% \\
        
        \rowcolor{lightgray}
        $\vips_6$ & 13 & 44.5\% & 55.5\% \\

        $\vips_{11}$ & 108 & 72.5\% & 27.5\% \\

        \approach & 56 & \textbf{74.5\%} & 25.5\% \\
        
        \bottomrule
    \end{tabular}
}
\end{table}

As demonstrated in \autoref{tab:classification-stats}, \approach's segmentation heuristic successfully fulfills its design objective of prioritizing component-like structures. On average, 74.5\% of the segments generated by \approach are subsequently classified as \textit{Components}. This represents a 2.8\% increase over the rate achieved by the best-performing \vips variant, while \approach concurrently produces, on average, 48.1\% fewer segments overall.

\subsection{Feature Inference and Test Generation Comparison (RQ3)}
\label{sec:eval-feature}
To evaluate the efficacy of \approach in E2E test generation, we benchmark it against \autoee. \autoee introduced the paradigm of feature-driven E2E testing and demonstrated superior performance compared to various traditional and earlier LLM-based E2E test generation techniques~\cite{alian2025feature}. To the best of our knowledge, \autoee remains the state-of-the-art method in this specific domain and serves as our primary baseline.

Our experimental setup utilizes the \eebench~\cite{alian2025feature} benchmark, used also by \autoee. This benchmark comprises eight open-source web applications of varying complexities, providing a standardized suite for evaluation. We adopt their proposed metric, \textit{feature coverage}, which quantifies the proportion of an application's features that are exercised by a generated test suite. This metric directly measures the effectiveness of feature-driven testing approaches, and we use it to evaluate the test suites produced by \approach.

\begin{table}
{\small
    \caption{Feature Inference of \approach on \eebench}
    \label{tab:inference-stats}
    \centering
    \begin{tabular}{lrrrrr}
        \toprule
        \textbf{App Name} & \textbf{Total} & \textbf{Correct} & \textbf{Precision} & \textbf{Recall} & \textbf{F1} \\
        \midrule

        \rowcolor{lightgray}
        PetClinic & 26 & 22 & 0.85 & 0.96 & 0.90 \\

        Conduit & 28 & 17 & 0.61 & 1.00 & 0.76 \\
        
        \rowcolor{lightgray}
        Taskcafe & 98 & 31 & 0.32 & 0.89 & 0.47 \\

        Dimeshift & 41 & 23 & 0.56 & 1.00 & 0.72 \\

        \rowcolor{lightgray}
        MantisBT & 143 & 26 & 0.18 & 0.93 & 0.30 \\

        EverTraduora & 64 & 37 & 0.58 & 0.90 & 0.70 \\

        \rowcolor{lightgray}
        Storefront & 47 & 13 & 0.28 & 1.00 & 0.43 \\

        Dashboard & 198 & 87 & 0.44 & 0.67 & 0.53 \\

        \midrule
        Total & 645 & 256 & 0.40 & 0.83 & 0.54 \\
        
        \bottomrule
    \end{tabular}
}
\end{table}

The feature coverage for the test suites generated by \approach is detailed in \autoref{tab:inference-stats}. For each \eebench subject, the table presents the total number of test cases generated by \approach, the count of those correctly covering actual application features, and the resulting precision, recall, and $F_1$-score. Aggregating performance across all applications, \approach achieves a global precision of 40\% (i.e., the proportion of all features inferred by \approach that were indeed correct) and a global recall of \textbf{\totalFeatureCoverage} (i.e., the proportion of all actual features across \eebench that \approach successfully identified). These aggregate scores yield a global $F_1$-score of 0.54.


\begin{figure}
    \centering
    \includegraphics[width=0.48\textwidth]{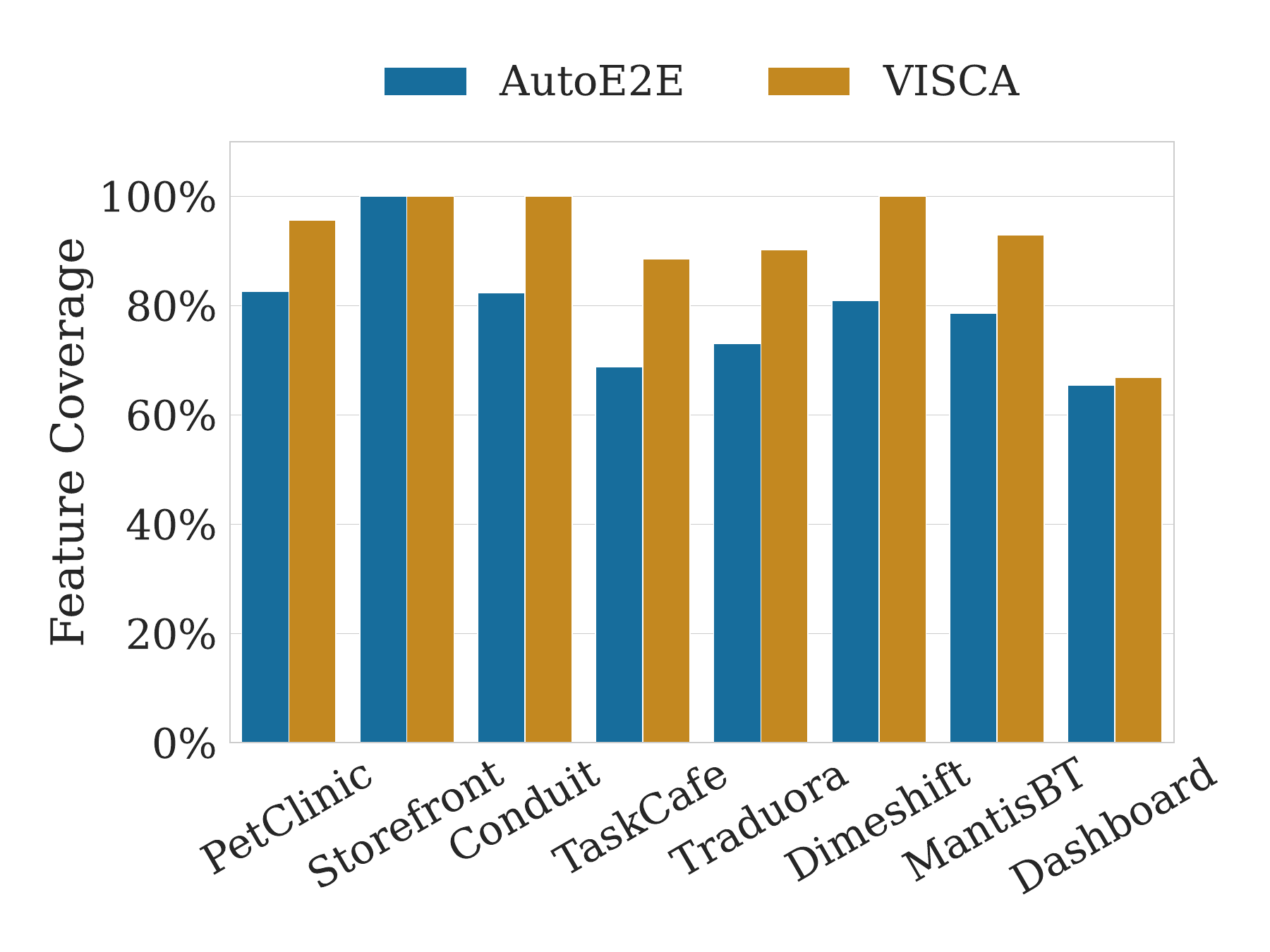}
    \caption{Feature Coverage of \approach and \autoee on \eebench}
    \label{fig:method-coverage}
    \reducespace
\end{figure}

Furthermore, a direct comparison of feature coverage achieved by \approach versus \autoee is presented in \autoref{fig:method-coverage}. Our method, \approach, attains an average feature coverage of \textbf{\featureCoverage}, exceeding the reported \autoeeCoverage for \autoee and demonstrating a \featureImprov improvement over this baseline. Notably, \approach shows superior or equal feature coverage compared to \autoee on every application in the benchmark, achieving perfect coverage on three of the eight applications. In terms of feature identification accuracy, \approach achieves an overall $F_1$-score of 0.54, compared to 0.62 for \autoee.

\section{Discussion}

\header{Segmentation Precision/Recall}
Our segmentation evaluation (Section \ref{sec:eval-segmentation}) indicates that \approach achieves higher precision than \vips at the cost of lower recall and $F_1$-score. We contend this trade-off is beneficial for our primary objective of component abstraction and feature inference. High precision ensures that the segments fed into our subsequent LLM-based transformation stage are coherent units, representing potential components, directly enhancing the quality of the abstraction and feature inference. The lower recall for \approach is a consequence of its heuristic design, producing a flat list of segments, contrasting with hierarchical segments that \vips and the ground-truth dataset exhibit. Therefore, for the goal of generating high-quality inputs for reliable component transformation and feature inference, \approach prioritizes segmentation precision over maximizing recall.

\header{Vanilla Prompting for Abstraction}
A simpler alternative to our multi-stage \approach might be direct, single-prompt LLM abstraction using a page's full HTML and visual rendering. However, this holistic method, while potentially viable for small pages, proves unreliable for complex commercial applications due to common LLM limitations such as overlooking details, hallucination, and exceeding context windows. For instance, a single-prompt \gemini attempt on Amazon's cart page (\autoref{fig:amazon-cart}) consistently failed to capture more fine-grained details in smaller sections (e.g., navigation, footer, list items) that \approach's structured process successfully abstracted. This disparity underscores the necessity of a methodical, multi-stage technique like \approach for robust component abstraction from real-world webpages.
\header{Threats to Validity}
It is important to acknowledge the potential threats to the validity of our work and the steps to mitigate them. One such threat lies in the representativeness of subjects for component abstraction evaluation in Section \ref{sec:eval-abstraction}. We have carefully examined all the possible webpages for this evaluation, and aimed to diversify and reduce the potential redundancy in the subjects that might skew our results by selecting fewer subjects from Material UI that showcase the same category of webpages.

Another threat lies in the use of LLMs in the evaluation of component abstraction (Section \ref{sec:eval-abstraction}). Using LLMs can be an unreliable solution, which is why we utilized the evaluation LLM in its reasoning mode to ensure a better understanding of the results, and independent authors validated the responses from the LLM to ensure its validity.

Finally, a temperature of 0 has been used for the LLM in our evaluations to ensure consistency. However, the results may still slightly vary in separate runs.

\section{Related Work}
\header{Segmentation}
Extensive research has been conducted on designing segmentation techniques for webpages. Many established techniques rely heavily on a single modality. For instance, DOM-based approaches~\cite{manabe2015extracting, rajkumar2012dynamic, vineel2009web, kang2010repetition} like the widely adopted VIPS algorithm \cite{cai2003vips} primarily analyze the HTML structure. While useful, DOM properties do not always align with visual presentation or semantic meaning \cite{bajammal2021page}, and the heuristics employed can be brittle or outdated, failing to capture how elements are contextually grouped from a user's perspective. Conversely, text-based methods \cite{kohlschutter2008densitometric, kolcz2007site} focus on linguistic analysis of textual content, often ignoring the crucial roles of visual layout, styling, and imagery in defining contextual relationships on the page. More recent visual approaches \cite{bajammal2021page, meier2017fully, cormer2017towards, chen1906mmdetection} leverage positioning and appearance but may lack grounding in the underlying semantic or structural information conveyed through text and the DOM.

\header{UI Code Generation}
Automated UI code generation aims to convert diverse design specifications into functional code, tackling the inefficiencies of manual development~\cite{kaluarachchi2023systematic}.
Earlier work in the field has ranged from utilizing heuristics to translate mockups into web components~\cite{bajammal2018generating} to utilizing neural networks and deep learning to translate screenshots of the target UI into code~\cite{beltramelli2018pix2code}.
Currently, LLMs and generative AI, including diffusion models~\cite{garg2025controllable}, are at the forefront, enabling sophisticated multimodal understanding and code synthesis~\cite{xiao2024prototype2code}, though achieving high semantic fidelity and maintainability in generated code remains a primary research focus~\cite{kaluarachchi2023systematic, xiao2024prototype2code}.

\header{LLMs for Test Generation}
Recent research in software testing increasingly emphasizes automation, with LLMs emerging as significant enablers across diverse testing domains. While a substantial body of work has leveraged LLMs for unit test generation~\cite{llmunittest, chen2024chatunitest, nan2025test, cheng2025rug, yin2024you}, and notable advancements have been made in automated API testing~\cite{kim2024multi} and accessibility testing~\cite{axnav, huq2024automated} using various techniques, a prominent recent trend is the focused application of LLMs to E2E test automation. Within this E2E context, current techniques are addressing specific sub-problems, such as enhancing semantic form testing~\cite{alian2024semantic}, or inferring application features to drive more comprehensive and fully automated E2E test generation~\cite{alian2025feature}.

In the specific domain of mobile application testing, LLMs are also being actively investigated. These efforts range from utilizing multi-stage LLM prompting to guide action selection for functional testing~\cite{liu2023chatting, liu2024make}, to deploying agentic frameworks where multiple LLM-driven agents interact with the application for thorough testing~\cite{wang2024xuat}. More recent work in mobile accessibility also explores leveraging LLMs for scenario-based evaluation~\cite{zhang2025scenario}.

\section{Conclusion}
In this work, we introduced \approach for transforming web applications into a component abstraction and utilizing the abstraction for automated E2E test generation. Our evaluations demonstrate that \approach achieves a \textbf{\featureCoverage} feature coverage in its generated test cases, outperforming the baseline by \textbf{\featureImprov}. Furthermore, the evaluation of our core component abstraction methodology reveals a high degree of fidelity, attaining an average structural match of \textbf{\structureMatchAverage} and an average content match of \textbf{\contentMatchAverage} between the original webpages and their generated representations. This high-quality abstraction not only validates the effectiveness of our transformation process but also underscores its strong potential for enhancing various downstream tasks, prominently including the improved E2E test generation demonstrated in this paper.


\bibliographystyle{IEEEtran}
\interlinepenalty=10000
\bibliography{references}

\begin{thebibliography}{10}
\providecommand{\url}[1]{#1}
\csname url@samestyle\endcsname
\providecommand{\newblock}{\relax}
\providecommand{\bibinfo}[2]{#2}
\providecommand{\BIBentrySTDinterwordspacing}{\spaceskip=0pt\relax}
\providecommand{\BIBentryALTinterwordstretchfactor}{4}
\providecommand{\BIBentryALTinterwordspacing}{\spaceskip=\fontdimen2\font plus
\BIBentryALTinterwordstretchfactor\fontdimen3\font minus \fontdimen4\font\relax}
\providecommand{\BIBforeignlanguage}[2]{{%
\expandafter\ifx\csname l@#1\endcsname\relax
\typeout{** WARNING: IEEEtran.bst: No hyphenation pattern has been}%
\typeout{** loaded for the language `#1'. Using the pattern for}%
\typeout{** the default language instead.}%
\else
\language=\csname l@#1\endcsname
\fi
#2}}
\providecommand{\BIBdecl}{\relax}
\BIBdecl

\bibitem{mesbah2012crawling}
A.~Mesbah, A.~Van~Deursen, and S.~Lenselink, ``Crawling ajax-based web applications through dynamic analysis of user interface state changes,'' \emph{ACM Transactions on the Web (TWEB)}, vol.~6, no.~1, pp. 1--30, 2012.

\bibitem{biagiola2019diversity}
M.~Biagiola, A.~Stocco, F.~Ricca, and P.~Tonella, ``Diversity-based web test generation,'' in \emph{Proceedings of the 2019 27th ACM Joint Meeting on European Software Engineering Conference and Symposium on the Foundations of Software Engineering}, 2019, pp. 142--153.

\bibitem{web-fragmentsRahul}
R.~K. Yandrapally and A.~Mesbah, ``{Fragment-Based Test Generation for Web Apps},'' \emph{IEEE Transactions on Software Engineering}, vol.~49, no.~3, pp. 1086--1101, 2023.

\bibitem{web-matteo-icst20}
M.~Biagiola, A.~Stocco, F.~Ricca, and P.~Tonella, ``{Dependency-Aware Web Test Generation},'' in \emph{2020 IEEE 13th International Conference on Software Testing, Validation and Verification (ICST)}, 2020, pp. 175--185.

\bibitem{web-chang2023reinforcement}
X.~Chang, Z.~Liang, Y.~Zhang, L.~Cui, Z.~Long, G.~Wu, Y.~Gao, W.~Chen, J.~Wei, and T.~Huang, ``{A Reinforcement Learning Approach to Generating Test Cases for Web Applications},'' in \emph{2023 IEEE/ACM International Conference on Automation of Software Test (AST)}, 2023, pp. 13--23.

\bibitem{nan2025test}
Z.~Nan, Z.~Guo, K.~Liu, and X.~Xia, ``Test intention guided llm-based unit test generation,'' in \emph{2025 IEEE/ACM 47th International Conference on Software Engineering (ICSE)}.\hskip 1em plus 0.5em minus 0.4em\relax IEEE Computer Society, 2025, pp. 779--779.

\bibitem{cheng2025rug}
X.~Cheng, F.~Sang, Y.~Zhai, X.~Zhang, and T.~Kim, ``Rug: Turbo llm for rust unit test generation,'' in \emph{2025 IEEE/ACM 47th International Conference on Software Engineering (ICSE)}.\hskip 1em plus 0.5em minus 0.4em\relax IEEE Computer Society, 2025, pp. 634--634.

\bibitem{yin2024you}
X.~Yin, C.~Ni, X.~Xu, and X.~Yang, ``What you see is what you get: Attention-based self-guided automatic unit test generation,'' \emph{arXiv preprint arXiv:2412.00828}, 2024.

\bibitem{kim2024multi}
M.~Kim, T.~Stennett, S.~Sinha, and A.~Orso, ``A multi-agent approach for rest api testing with semantic graphs and llm-driven inputs,'' \emph{arXiv preprint arXiv:2411.07098}, 2024.

\bibitem{huq2024automated}
S.~F. Huq, M.~Tafreshipour, K.~Kalcevich, and S.~Malek, ``Automated generation of accessibility test reports from recorded user transcripts,'' in \emph{2025 IEEE/ACM 47th International Conference on Software Engineering (ICSE)}.\hskip 1em plus 0.5em minus 0.4em\relax IEEE Computer Society, 2024, pp. 534--546.

\bibitem{alian2025feature}
P.~Alian, N.~Nashid, M.~Shahbandeh, T.~Shabani, and A.~Mesbah, ``Feature-driven end-to-end test generation,'' in \emph{2025 IEEE/ACM 47th International Conference on Software Engineering (ICSE)}.\hskip 1em plus 0.5em minus 0.4em\relax IEEE Computer Society, 2025, pp. 678--678.

\bibitem{react}
``React,'' \url{https://react.dev/}, 2025, accessed: 2025-04-10.

\bibitem{vuejs}
``Vuejs,'' \url{https://vuejs.org/}, 2025, accessed: 2025-04-10.

\bibitem{svelte}
``Svelte,'' \url{https://svelte.dev/}, 2025, accessed: 2025-04-10.

\bibitem{cai2003vips}
D.~Cai, S.~Yu, J.-R. Wen, and W.-Y. Ma, ``Vips: a vision-based page segmentation algorithm,'' 2003.

\bibitem{vineel2009web}
G.~Vineel, ``Web page dom node characterization and its application to page segmentation,'' in \emph{2009 IEEE International Conference on Internet Multimedia Services Architecture and Applications (IMSAA)}.\hskip 1em plus 0.5em minus 0.4em\relax IEEE, 2009, pp. 1--6.

\bibitem{kang2010repetition}
J.~Kang, J.~Yang, and J.~Choi, ``Repetition-based web page segmentation by detecting tag patterns for small-screen devices,'' \emph{IEEE Transactions on Consumer Electronics}, vol.~56, no.~2, pp. 980--986, 2010.

\bibitem{kolcz2007site}
A.~Ko{\l}cz and W.-t. Yih, ``Site-independent template-block detection,'' in \emph{European Conference on Principles of Data Mining and Knowledge Discovery}.\hskip 1em plus 0.5em minus 0.4em\relax Springer, 2007, pp. 152--163.

\bibitem{meier2017fully}
B.~Meier, T.~Stadelmann, J.~Stampfli, M.~Arnold, and M.~Cieliebak, ``Fully convolutional neural networks for newspaper article segmentation,'' in \emph{2017 14th IAPR International conference on document analysis and recognition (ICDAR)}, vol.~1.\hskip 1em plus 0.5em minus 0.4em\relax IEEE, 2017, pp. 414--419.

\bibitem{cormer2017towards}
M.~Cormer, R.~Mann, K.~Moffatt, and R.~Cohen, ``Towards an improved vision-based web page segmentation algorithm,'' in \emph{2017 14th Conference on Computer and Robot Vision (CRV)}.\hskip 1em plus 0.5em minus 0.4em\relax IEEE, 2017, pp. 345--352.

\bibitem{chen1906mmdetection}
K.~Chen, J.~Wang, J.~Pang, Y.~Cao, Y.~Xiong, X.~Li, S.~Sun, W.~Feng, Z.~Liu, J.~Xu \emph{et~al.}, ``Mmdetection: Open mmlab detection toolbox and benchmark. arxiv 2019,'' \emph{arXiv preprint arXiv:1906.07155}, 1906.

\bibitem{bajammal2021page}
M.~Bajammal and A.~Mesbah, ``Page segmentation using visual adjacency analysis,'' \emph{arXiv preprint arXiv:2112.11975}, 2021.

\bibitem{kiesel2021empirical}
J.~Kiesel, L.~Meyer, F.~Kneist, B.~Stein, and M.~Potthast, ``An empirical comparison of web page segmentation algorithms,'' in \emph{European Conference on Information Retrieval}.\hskip 1em plus 0.5em minus 0.4em\relax Springer, 2021, pp. 62--74.

\bibitem{bootstrap}
``Bootstrap,'' \url{https://getbootstrap.com/}, 2025, accessed: 2025-04-10.

\bibitem{mui}
``Material ui,'' \url{https://mui.com/}, 2025, accessed: 2025-04-10.

\bibitem{antd}
``Ant design,'' \url{https://ant.design/}, 2025, accessed: 2025-04-10.

\bibitem{shadcn}
``Shadcn,'' \url{https://ui.shadcn.com/}, 2025, accessed: 2025-04-10.

\bibitem{chakra}
``Chakra ui,'' \url{https://chakra-ui.com/}, 2025, accessed: 2025-04-10.

\bibitem{bajammal2018generating}
M.~Bajammal, D.~Mazinanian, and A.~Mesbah, ``Generating reusable web components from mockups,'' in \emph{Proceedings of the 33rd ACM/IEEE International Conference on Automated Software Engineering}, 2018, pp. 601--611.

\bibitem{zhang1989simple}
K.~Zhang and D.~Shasha, ``Simple fast algorithms for the editing distance between trees and related problems,'' \emph{SIAM journal on computing}, vol.~18, no.~6, pp. 1245--1262, 1989.

\bibitem{mongodb}
``Mongodb atlas,'' \url{https://www.mongodb.com}, 2024, accessed: 2025-04-10.

\bibitem{selenium}
``Selenium,'' \url{https://www.selenium.dev}, 2024, accessed: 2025-04-10.

\bibitem{kreuzer2013quantitative}
R.~Kreuzer, ``A quantitative comparison of semantic web page segmentation algorithms,'' Master's thesis, 2013.

\bibitem{kiesel2020web}
J.~Kiesel, F.~Kneist, L.~Meyer, K.~Komlossy, B.~Stein, and M.~Potthast, ``Web page segmentation revisited: evaluation framework and dataset,'' in \emph{Proceedings of the 29th ACM International Conference on Information \& Knowledge Management}, 2020, pp. 3047--3054.

\bibitem{amigo2009comparison}
E.~Amig{\'o}, J.~Gonzalo, J.~Artiles, and F.~Verdejo, ``A comparison of extrinsic clustering evaluation metrics based on formal constraints,'' \emph{Information retrieval}, vol.~12, pp. 461--486, 2009.

\bibitem{manabe2015extracting}
T.~Manabe and K.~Tajima, ``Extracting logical hierarchical structure of html documents based on headings,'' \emph{Proceedings of the VLDB Endowment}, vol.~8, no.~12, pp. 1606--1617, 2015.

\bibitem{rajkumar2012dynamic}
K.~Rajkumar and V.~Kalaivani, ``Dynamic web page segmentation based on detecting reappearance and layout of tag patterns for small screen devices,'' in \emph{2012 International Conference on Recent Trends in Information Technology}.\hskip 1em plus 0.5em minus 0.4em\relax IEEE, 2012, pp. 508--513.

\bibitem{kohlschutter2008densitometric}
C.~Kohlsch{\"u}tter and W.~Nejdl, ``A densitometric approach to web page segmentation,'' in \emph{Proceedings of the 17th ACM conference on Information and knowledge management}, 2008, pp. 1173--1182.

\bibitem{kaluarachchi2023systematic}
T.~Kaluarachchi and M.~Wickramasinghe, ``A systematic literature review on automatic website generation,'' \emph{Journal of Computer Languages}, vol.~75, p. 101202, 2023.

\bibitem{beltramelli2018pix2code}
T.~Beltramelli, ``pix2code: Generating code from a graphical user interface screenshot,'' in \emph{Proceedings of the ACM SIGCHI symposium on engineering interactive computing systems}, 2018, pp. 1--6.

\bibitem{garg2025controllable}
A.~Garg, Y.~Jiang, and A.~Oulasvirta, ``Controllable gui exploration,'' \emph{arXiv preprint arXiv:2502.03330}, 2025.

\bibitem{xiao2024prototype2code}
S.~Xiao, Y.~Chen, J.~Li, L.~Chen, L.~Sun, and T.~Zhou, ``Prototype2code: End-to-end front-end code generation from ui design prototypes,'' in \emph{International Design Engineering Technical Conferences and Computers and Information in Engineering Conference}, vol. 88353.\hskip 1em plus 0.5em minus 0.4em\relax American Society of Mechanical Engineers, 2024, p. V02BT02A038.

\bibitem{llmunittest}
M.~Schäfer, S.~Nadi, A.~Eghbali, and F.~Tip, ``An empirical evaluation of using large language models for automated unit test generation,'' \emph{IEEE Transactions on Software Engineering}, vol.~50, no.~1, pp. 85--105, 2024.

\bibitem{chen2024chatunitest}
Y.~Chen, Z.~Hu, C.~Zhi, J.~Han, S.~Deng, and J.~Yin, ``Chatunitest: A framework for llm-based test generation,'' in \emph{Companion Proceedings of the 32nd ACM International Conference on the Foundations of Software Engineering}, 2024, pp. 572--576.

\bibitem{axnav}
\BIBentryALTinterwordspacing
M.~Taeb, A.~Swearngin, E.~Schoop, R.~Cheng, Y.~Jiang, and J.~Nichols, ``Axnav: Replaying accessibility tests from natural language,'' in \emph{Proceedings of the CHI Conference on Human Factors in Computing Systems}, ser. CHI '24.\hskip 1em plus 0.5em minus 0.4em\relax New York, NY, USA: Association for Computing Machinery, 2024. [Online]. Available: \url{https://doi.org/10.1145/3613904.3642777}
\BIBentrySTDinterwordspacing

\bibitem{alian2024semantic}
P.~Alian, N.~Nashid, M.~Shahbandeh, and A.~Mesbah, ``Semantic constraint inference for web form test generation,'' in \emph{Proceedings of the 33rd ACM SIGSOFT International Symposium on Software Testing and Analysis}, 2024, pp. 932--944.

\bibitem{liu2023chatting}
Z.~Liu, C.~Chen, J.~Wang, M.~Chen, B.~Wu, X.~Che, D.~Wang, and Q.~Wang, ``Chatting with gpt-3 for zero-shot human-like mobile automated gui testing. arxiv 2023,'' \emph{arXiv preprint arXiv:2305.09434}, 2023.

\bibitem{liu2024make}
------, ``Make llm a testing expert: Bringing human-like interaction to mobile gui testing via functionality-aware decisions,'' in \emph{Proceedings of the IEEE/ACM 46th International Conference on Software Engineering}, 2024, pp. 1--13.

\bibitem{wang2024xuat}
Z.~Wang, W.~Wang, Z.~Li, L.~Wang, C.~Yi, X.~Xu, L.~Cao, H.~Su, S.~Chen, and J.~Zhou, ``Xuat-copilot: Multi-agent collaborative system for automated user acceptance testing with large language model,'' \emph{arXiv preprint arXiv:2401.02705}, 2024.

\bibitem{zhang2025scenario}
Y.~Zhang, S.~Chen, X.~Xie, Z.~Liu, and L.~Fan, ``Scenario-driven and context-aware automated accessibility testing for android apps,'' in \emph{2025 IEEE/ACM 47th International Conference on Software Engineering (ICSE)}.\hskip 1em plus 0.5em minus 0.4em\relax IEEE Computer Society, 2025, pp. 630--630.

\end{thebibliography}

\end{document}